\documentclass[fleqn,twoside]{article}
\usepackage{espcrc2}

\usepackage[latin1]{inputenc}
\usepackage[english]{babel}
\usepackage{graphicx}
\usepackage{pgf,pgfarrows,pgfnodes,pgfautomata,pgfheaps,pgfshade}
\usepackage{pstricks,pst-node} 
\usepackage{cite}
\usepackage{amssymb}
\usepackage{subfigure}
\usepackage{xspace}
\usepackage{amsmath}
\usepackage{units}

\newrgbcolor{verydarkgray}{0.2 0.2 0.2}
\newrgbcolor{darkgray}{0.3 0.3 0.3}
\newrgbcolor{lightgray}{0.5 0.5 0.5} 
\newrgbcolor{verylightgray}{0.9 0.9 0.9}
\newrgbcolor{lightgreen}{0.85 1.0 0.8}
\newrgbcolor{lightpurple}{0.95 0.88 1.0}
\newrgbcolor{lightblue}{0.85 0.95 1.0}
\newrgbcolor{lightred}{1.0 0.9 0.85}
\newrgbcolor{lightorange}{1.0 1.0 0.85}
\newrgbcolor{verylightblue}{0.93 0.93 1.0}
\newrgbcolor{darkblue}{0.3 0.3 0.7}
\newrgbcolor{darkgreen}{0.1 0.7 0.1}
\newrgbcolor{darkred}{.7 0.3 0.3}
\newrgbcolor{darkorange}{0.65 0.25 0.15}

\title{Proton-Air Cross Section and Extensive Air Showers}

\author{Ralf~Ulrich\address[FZK]{Karlsruhe Institute of Technology
    (KIT)\\
    Institut f\"ur Kernphysik, P.O. Box 3640, 76021 Karlsruhe, Germany\\
    \emph{(KIT is the cooperation of University
    Karlsruhe and Forschungszentrum
    Karlsruhe)}}%
  , Ralph~Engel\addressmark[FZK],
  Steffen~M\"uller\addressmark[FZK],
  Fabian~Sch{\"u}ssler\addressmark[FZK] and
  Michael~Unger\addressmark[FZK] }

\begin{document}

\begin{abstract}
  Hadronic cross sections at ultra-high energy have a significant
  impact on the development of extensive air shower
  cascades. Therefore the interpretation of air shower data depends
  critically on hadronic interaction models that extrapolate the cross
  section from accelerator measurements to the highest cosmic ray
  energies. We discuss how extreme scenarios of cross section
  extrapolations can affect the interpretation of air shower data.  We
  find that the theoretical uncertainty of the extrapolated proton-air cross
  section at ultra-high energies is much larger than suggested
  by the existing spread of available Monte Carlo model predictions. The impact on
  the depth of the shower maximum is demonstrated.
\end{abstract}

\maketitle

%
\section{Introduction}

The inherent uncertainties of the modeling of hadronic interactions at
ultra-high energies are still preventing an unambiguous analysis of
existing cosmic ray data. They introduce large and difficult to
quantify systematic uncertainties on many cosmic ray analyses. This is
a well known problem for the primary energy spectrum as derived from
Monte Carlo calibrated surface detector
data~\cite{Nagano:1992jz,Takeda:1998ps,Ave:2001hq,Aglietta:2004np,%
  Antoni:2005wq,Amenomori:2008jb}, as well as for the determination of
the primary mass composition of high energy cosmic
rays~\cite{Dawson:1998kk,Ave:2002gc,Dova:2003an,Abbasi:2004nz,Antoni:2005wq,Unger:2007mc}.
The model-dependence of the interpretation of air shower data is
mainly related to the differing predictions of the number of muons in
air showers and the depth of the shower maximum~\cite{Heck08CORSIKASchool}.

It is unclear whether the existing differences in model
predictions can be used to estimate the current theoretical systematic
uncertainties caused by inaccurate modeling of hadronic interactions in
air showers. It is possible that existing
model-differences are
\begin{enumerate}
\item[(A)] \emph{overestimating} the current systematic uncertainties.
  Better understanding of hadronic particle production and new data
  from accelerators allow us to update interaction models to obtain a
  more realistic description of particle production. Not all models
  are updated regularly and the quality of data description differs
  between the models. 
\item[(B)] \emph{underestimating} the current
  systematic uncertainties, because
  the existing models are by far not sampling the full phase space of
  possible interaction scenarios and parameters. Moreover, new physics at higher energies may
  be unknown and is thus even missing in current approaches.
\end{enumerate}

Of the frequently used models, there seem to be more \emph{sophisticated} and possibly better
models,
e.g.\ \textsc{QGSJet~II}~\cite{Ostapchenko:2004ss,Ostapchenko:2006vr}
and \textsc{Epos}~\cite{Werner:2007vd}, compared to others that are
\emph{older} and more simplistic or specialized,
e.g.\ \textsc{QGSJet~01}~\cite{Kalmykov:1993qe,Kalmykov:1989br} and
\textsc{Sibyll~2.1}~\cite{Engel:1999db,Fletcher:1994bd}. Furthermore it
must be kept in mind that many other models
are considered as \emph{outdated} or just not \emph{fashionable} like
\textsc{neXus}~\cite{Drescher:2000ha,Hladik:2001zy},
\textsc{HDMP}~\cite{Capdevielle:1989ht}, \textsc{DPMJET}~\cite{Ranft:1994fd},
\textsc{DTUJET}~\cite{Bopp:1994cg},
\textsc{VENUS}~\cite{Werner:1993uh} etc.\,.  This
classification is very subjective; Different communities make
differing choices.

\begin{figure}[t!]
  \centering
  \includegraphics[width=.99\linewidth]{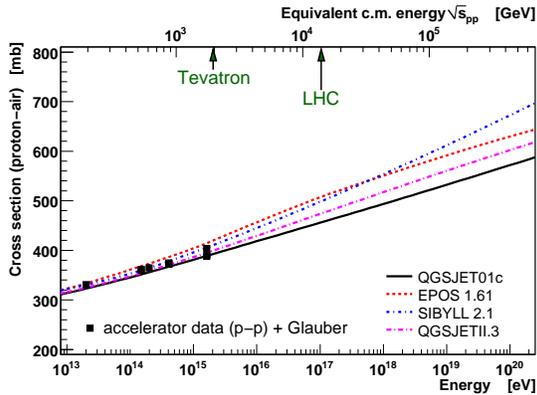}
  \vspace*{-.9cm}
  \caption{Extrapolation of the proton-air production cross section
    from accelerator measurements (see \protect\cite{Engel:2000pb} for
    references) to cosmic ray energies. The accelerator measurements (p-p
    and p-$\bar{\rm p}$) are converted to p-air within the framework
    of the respective interaction model.}
  \label{fig:sigma}
\end{figure}
The current situation is characterized by the fact that none of the
models is able to consistently describe cosmic ray data, and the
predictions by \textsc{Sibyll~2.1} and \textsc{QGSJet~01} are not
objectively worse than the ones by \textsc{QGSJetII} and
\textsc{Epos}. Over the years model predictions and extrapolations
have become more alike even though there is no theory for calculating
e.g. such cross sections from first principles. This is demonstrated
in Fig.~\ref{fig:sigma} by comparing the predictions on the total
proton-air cross section for secondary particle production. On the
other hand, recent activity in model development within the
\textsc{Epos} event generator clearly demonstrated that there exists a
relatively large freedom to tune model predictions at ultra-high
energies~\cite{Pierog:2007x1}. All this is indicating that the current
situation corresponds to option~(B).

The aim of this work is to start from existing accelerator
measurements and investigate the importance of different cross section
extrapolations. 
The theoretical formalism to calculate cross sections of
hadronic projectiles with air is described and the available
accelerator data are discussed. Based on these considerations
different extrapolations of the existing data to ultra-high energy
are considered and the resulting impact on the development and, thus,
interpretation of cosmic ray induced air showers
are discussed.

%
\section{Glauber Formalism}
\label{sec:Glauber}

For the development of extensive air showers the cross sections of
hadronic projectiles with air at center of mass energies from GeV to
several hundreds of TeV are relevant. These cross sections have to be
calculated from models based on data for nucleon-nucleon and
meson-nucleon scattering that has been studied at accelerators.

It was realized by Glauber~\cite{Glauber:1955qq,Glauber:1970jm} that
the scattering amplitude in the impact parameter space can be treated
in analogy to diffraction in optics by using the phase shift relation
\begin{equation}
  a(s,\,b)=1-e^{{\rm i}\chi(s,\,b)} \,,
\end{equation}
with the Eikonal\footnote{From the Greek word
  $\epsilon\iota\kappa\omega\nu$ with the meaning of image.}  function
$\chi$. Within this picture of neglecting the recoil due to the individual
nucleon-nucleon interactions, multiple
scattering can be written as
\begin{align}
  \label{eqn::phaseshift}
  a_{hA}(s,\,b)&=1-e^{{\rm i}\chi_{\rm mult}(s,\,b)}\nonumber\\ 
&=1-\prod\limits_j\,\left(1-a_j(s,\,b_j)\right)
  \,,
\end{align}
whereas the overall phase shift $\chi_{\rm mult}$ corresponds to the
sum of the individual phase shifts $\chi_j$ for each scattering
process $j$.

\begin{figure}[t!]
  \centering
  \includegraphics[width=\linewidth]{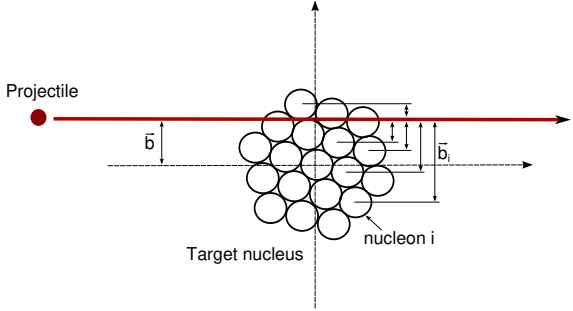}
  \vspace*{-1cm}
  \caption{A projectile particle hitting a target nucleus with an
    impact parameter $\vec{b}$. The relative impact parameter $\vec{b}_j$
    for each of the nucleons governs the scattering of the
    projectile from the individual nucleon $i$.}
  \label{f:Glauber}
\end{figure}

Figure~\ref{f:Glauber} illustrates the interaction of a projectile
particle with a nucleus. According to Eq.~(\ref{eqn::phaseshift}) the
projectile scatters separately with each of the nucleons $j$ of the
nucleus.
The resulting total and elastic cross section are given by
\begin{eqnarray}
\sigma^{\rm hA}_{\rm tot} &=& 2 \Re e\int \Gamma_{hA}(\vec b) {\rm d}^2 b,
\nonumber\\
\sigma^{\rm hA}_{\rm ela} &=& \int \left|  \Gamma_{hA}(\vec b) \right|^2
{\rm d}^2 b,
\label{eqn::glauber}
\end{eqnarray}
with
\begin{equation}
 \Gamma_{\rm hA}(\vec b)
 =  1 - \prod_{j=1}^{A} \left[ 1 - \int a_j(\vec b
  -\vec b_j)\rho_j(\vec r_j) {\rm d}^3 r_j\right].
\end{equation}
For clarity of the expressions we have applied the simplification of
non-correlated nucleons
\begin{equation}
\psi^\star(\vec r_1 \dots \vec r_A) \psi(\vec r_1 \dots \vec r_A)
=  \prod_{j=1}^A \rho_j(\vec r_j),
\end{equation}
where $\psi$ denotes the wave function of the nucleus with $\vec{r}_i$
being the coordinates of the individual nucleons $j=1 \dots A$. The
impact parameters of the nucleons are denoted by $\vec{b}_j$ and that
of the cosmic ray hadron by $\vec{b}$. The single nucleon
density is normalized to $\int \rho_j(\vec r_j) {\rm d}^3 r_j = 1$.
The sum of the cross sections for elastic and quasi-elastic scattering
is given by
\begin{eqnarray}
\sigma^{\rm hA}_{\rm ela} + \sigma^{\rm hA}_{\rm qel} =
\int  {\rm d}^2b \left| 1 - \prod_{j=1}^{A} \left[ 1 - a_j(\vec b
  -\vec s_j)\right] \right|^2
\nonumber\\
\times \left( \prod_{j=1}^A\rho_j(\vec r_j) {\rm d}^3r_j\right).
\end{eqnarray}

Knowing  the nucleon-nucleon impact parameter amplitudes
$a_j(s,\,\vec{b} -\vec{b}_j)$
allows one to calculate the multiple scattering amplitude and hence
the total, elastic and quasi-elastic cross sections.  

Based on the optical theorem the nucleon-nucleon scattering amplitude
is typically parametrized as
\begin{align}
  \label{eqn::amplitudeIII}
  a_j(s,\,b_j)=\frac{\left(1+\rho(s)\right) \sigma^{\rm pp}_{\rm
      tot}(s)}{4\pi B_{\rm ela}(s)} \, {\rm
    e}^{-\vec{b_j}^2/(2\,B_{\rm ela}(s))} ,
\end{align}
where $\rho$ is the ratio of the real to imaginary part of the forward
scattering amplitude \cite{Block:1984ru}, and $B_{\rm
  ela}$ is the elastic slope parameter defined by
${\rm d}\sigma_{\rm ela}/{\rm d}t|_{t=0}\propto{\rm e}^{-B_{\rm
    ela}|t|}$.

It is the combination of expressions (\ref{eqn::glauber}) and
(\ref{eqn::amplitudeIII}) that finally makes the connection between
nucleon-nucleon and nucleon-nucleus scattering.  With these
formulations it is possible to calculate the total, elastic and
quasi-elastic cross section for nucleon-nucleus scattering just from
the parameters $B_{\rm ela}$, $\rho$, $\sigma^{\rm pp}_{\rm tot}$ and
the possible configurations of the nucleus. 


It can be shown that the Glauber approximation follows from Gribov's
Regge calculus if inelastic intermediate states are neglected
\cite{Weis:1976er}. The inclusion of intermediate states leads to
inelastic screening corrections. \textsc{Sibyll~2.1} is a model that is based on
the classical Glauber approximation. Inelastic screening corrections
are implemented in \textsc{QGSJet~II} within Gribov's Reggeon calculus. The
cross section calculation in \textsc{Epos} is accounting for high parton
density effects with a parametrization of the expected screening
effects. There are other models of amplitudes and corresponding
schemes of describing multiple scattering that are not available
in any event generator, see for example \cite{Troshin93a}.

%
\section{Accelerator Data and their Extrapolation to Cosmic Ray Energies}
\begin{figure*}[bt!]
  \centering 
  \subfigure[$\sigma^{\rm pp}_{\rm tot}$]{\includegraphics[width=.32\linewidth]{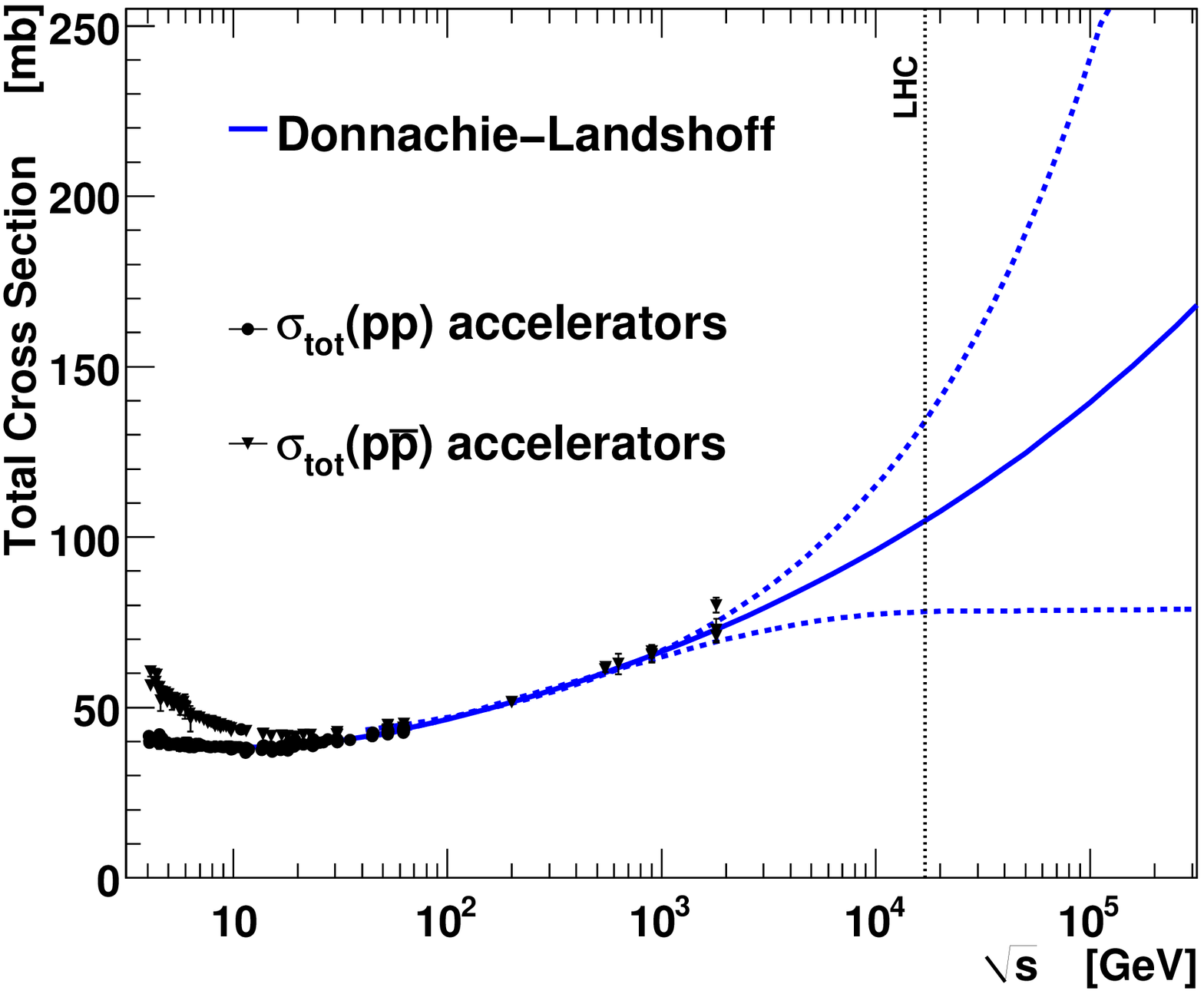}}
  \subfigure[$B_{\rm ela}$]{\includegraphics[width=.32\linewidth]{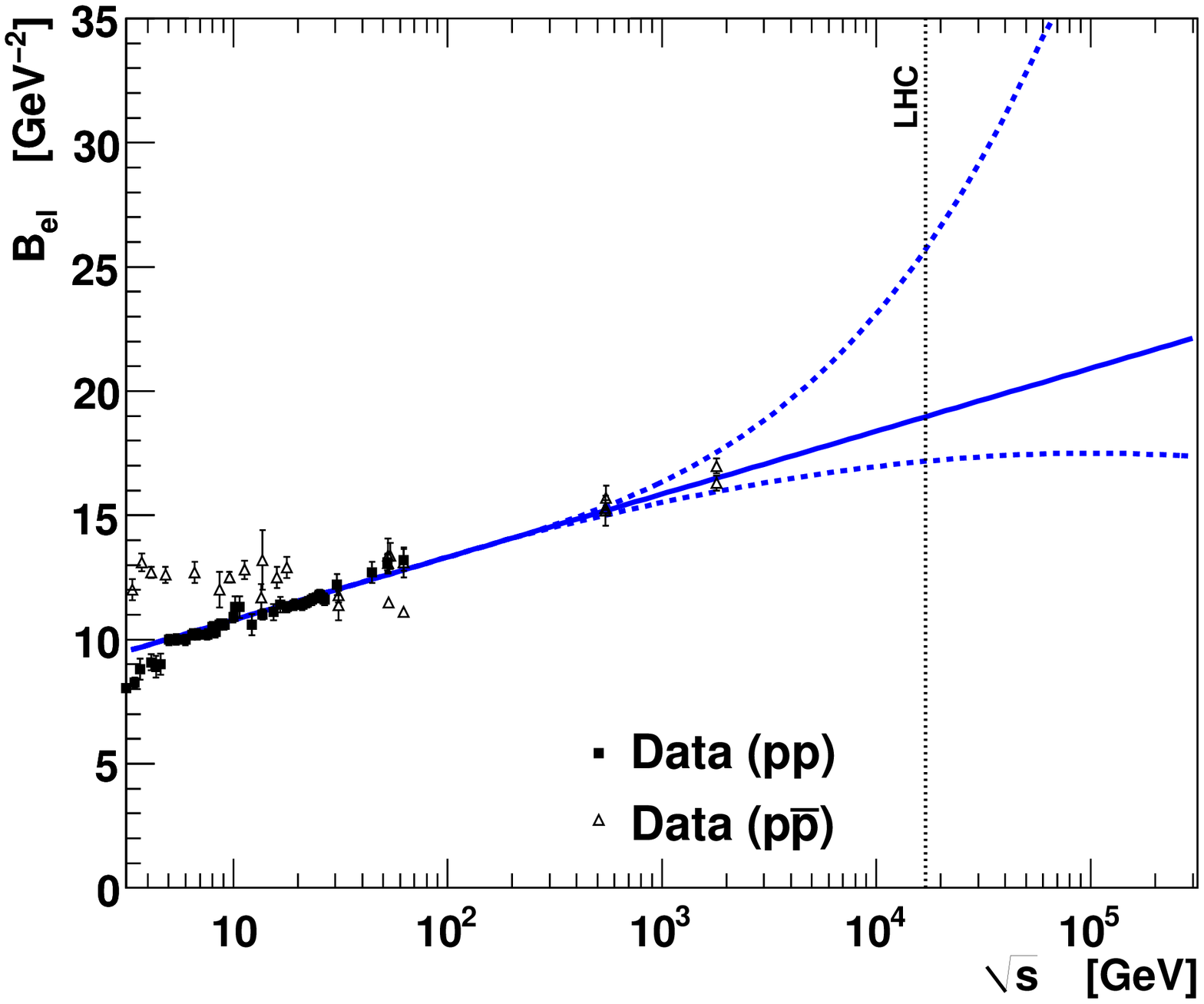}}
  \subfigure[$\rho$]{\includegraphics[width=.32\linewidth]{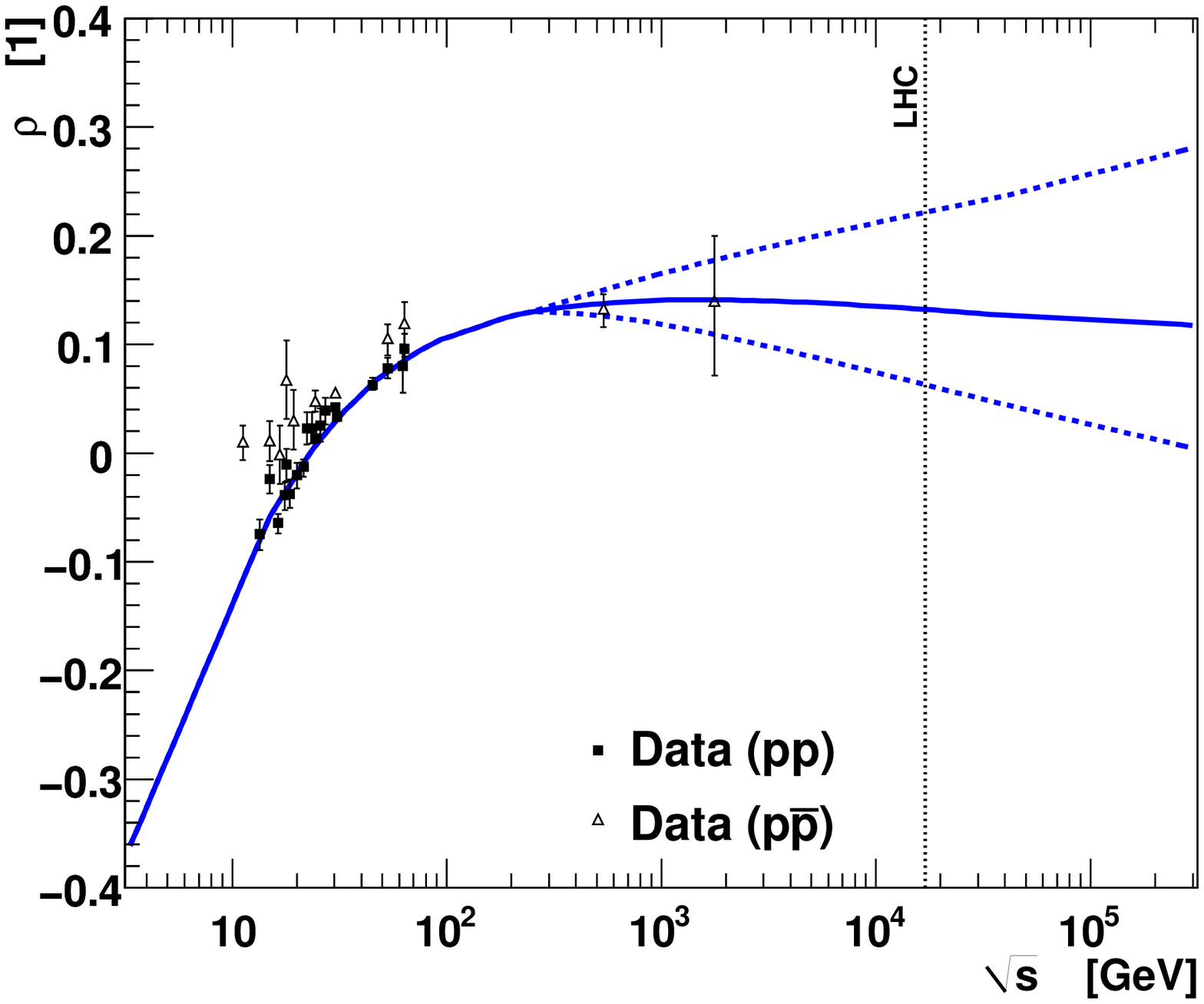}}
  \vspace*{-1cm}
  \caption{Compilation of accelerator data of $\sigma^{\rm
      pp}_{\rm tot}$, $B_{\rm ela}$ and
    $\rho$~\protect\cite{Engel:2000pb}. The central line denotes the
    conventional extrapolation of these data to high energy. The upper
    and lower lines indicate a set of possible extreme extrapolations.
    In the left plot the conventional model is the soft pomeron
    parametrization by
    Donnachie and Landshoff~\protect\cite{Donnachie:1992ny}, while the
    lower extreme is by Pancheri et al.~\protect\cite{Pancheri:2007rv}
    and the upper extreme is the two-pomeron model of
    Landshoff~\protect\cite{Landshoff:2007uk,Landshoff:2009wt}.}
  \label{fig:acceleratorData}
\end{figure*}
Up to $\sqrt{s}\unit[\sim2\times10^3]{GeV}$ measurements of $\rho$,
$B_{\rm ela}$ and $\sigma^{\rm pp}_{\rm tot}$ are available from
accelerator experiments~(c.f.~\cite{Engel:2000pb} and references
therein). The experimental precision is best at lower energies. The
extrapolation of these data to higher energies, however, is
model-dependent and subject to
uncertainties~(e.g.~\cite{Block:2006hy,Pancheri:2007rv,Landshoff:2009wt}). For the
application in cosmic ray physics at least two orders of magnitude in
energy in the center of mass system have to be extrapolated. In
Figure~\ref{fig:acceleratorData} the resulting magnitude of
uncertainty at cosmic ray energies is demonstrated. While all
extrapolations of $\rho$ and $B_{\rm ela}$ (c.f.
Fig.~\ref{fig:acceleratorData}, (b)+(c)) are just phenomenological parametrizations and thus
not based on underlying physics. The extrapolation of $\sigma_{\rm
  tot}^{\rm pp}$ (see Fig.~\ref{fig:acceleratorData}, (a)) is taken
from published work; The conventional model is the soft pomeron parametrization by
Donnachie and Landshoff~\cite{Donnachie:1992ny}, while the lower extreme
is a model by Pancheri et al.~\cite{Pancheri:2007rv} and the upper
extreme the two-pomeron model by
Landshoff~\cite{Landshoff:2007uk,Landshoff:2009wt}.

%
\section{Resulting Proton-Air Cross Section}
\begin{figure*}[t!]
  \centering
  \includegraphics[width=.65\linewidth]{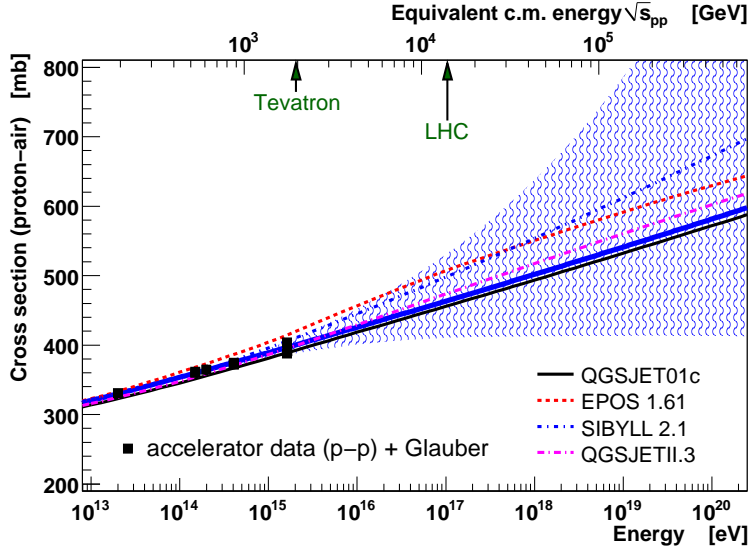}
  \vspace*{-1cm}
  \caption{Uncertainty of the extrapolation of the proton-air cross
    section, $\sigma_{\rm prod}^{\rm p-air}$, from accelerator to
    cosmic ray energies.}
  \label{fig:sigmaExt}
\end{figure*}

Combining the extrapolations given in Fig.~\ref{fig:acceleratorData}
with Glauber theory it is possible to calculate the proton-air cross
section, which is important for air shower development. Typically only
the \emph{production} cross section
\begin{equation}
  \sigma_{\rm prod}=\sigma_{\rm tot}-\sigma_{\rm ela}-\sigma_{\rm qel}
\end{equation}
 is quoted in the context of extensive air
 showers~\cite{Nikolaev:1993mc,Engel:1998pw}, since interactions with
 no new particle production are not relevant to the
 development of air showers.

The results are shown in Fig.~\ref{fig:sigmaExt}.  Already at
$\unit[10^{18}]{eV}$ the uncertainty band is significantly larger than
the range covered by all available interaction models. Around
$\unit[10^{19}]{eV}$ the relative uncertainty reaches up to \unit[50]{\%}.

%
\section{Impact on the Interpretation of Cosmic Ray Data}
\begin{figure*}[t!]
  \centering
  \includegraphics[width=.49\textwidth]{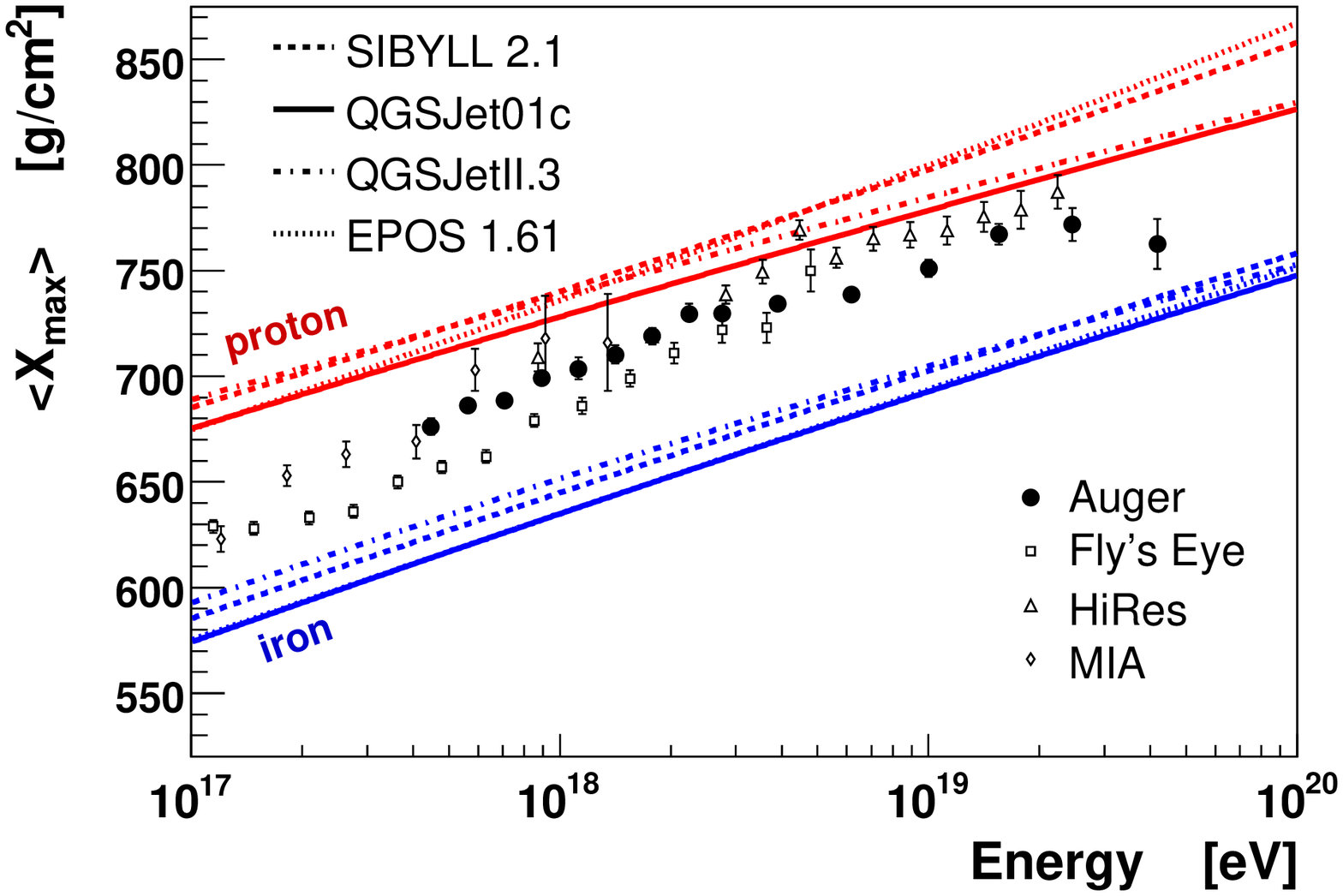}
  \includegraphics[width=.49\textwidth]{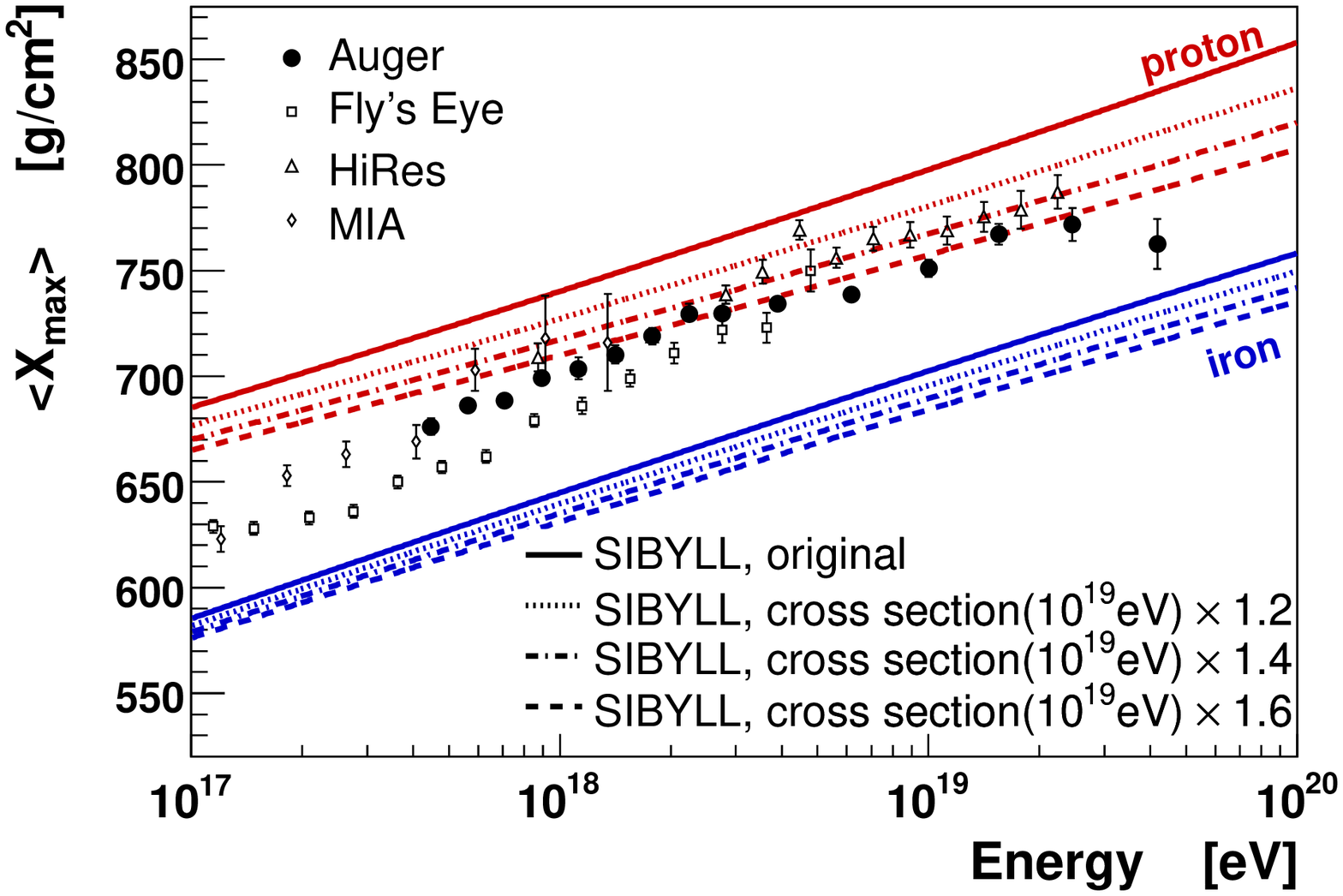}
  \caption{Left panel: Measurements of the mean depth of the shower
    maximum~\protect\cite{Bird:1993yi,AbuZayyad:2000ay,Abbasi:2004nz,Unger:2007mc}
    compared to air shower simulations for different primary particles
    and interaction models. Right panel: Same data compared to air
    shower simulations using \textsc{Sibyll} and a modified cross section
    extrapolation~\protect\cite{Ulrich:2007xv,Ulrich:2009xxx}.}
  \label{fig:xmax}
\end{figure*}

The choice of the extrapolation of the proton-air cross section
(Fig.~\ref{fig:sigmaExt}) has a strong impact on the predicted speed
of shower development and thus on the depth of the shower
maximum~\cite{Ulrich:2009xxx}. The typical interpretation of $\langle
X_{\rm max}\rangle$ data in terms of a mixed mass composition at high
energy (Fig.~\ref{fig:xmax}, left) has to be revised if a different
cross section extrapolation is used. As shown in
Fig.~\ref{fig:xmax}~(right), the data could also be explained with a
cross section that is increased by $\unit[f_{19 }= 40 - 60]{\%}$ at
$\unit[10^{19}]{eV}$ in combination with very light cosmic ray
primaries. Similar results have been obtained independently in
Ref.~\cite{Wibig:2008ji}.

In our calculations all hadronic interaction cross sections are
increased by a factor that depends logarithmically on energy
\begin{equation}
f(E) = 1 + (f_{19} -1) \frac{\ln (E/10^{15}\,{\rm eV})}{\ln (10^{19}\,{\rm eV}/10^{15}\,{\rm eV})},
\label{eq:rescalingFactor}
\end{equation}
for $E> 10^{15}$\,eV and $f(E) = 1$ otherwise.

%
\section{Summary}

It is argued that the uncertainties of the extrapolation of hadronic
cross sections to cosmic ray energies might be underestimated if only
commonly used models are considered. The true uncertainty could be
much larger than the one suggested by the spread of the current
predictions of hadronic interaction models.

Since longitudinal air shower development depends sensitively on
hadronic cross sections, predictions for standard observables like the
depth of the shower maximum, $X_{\rm max}$, are strongly affected by
these uncertainties. It is shown that the measured mean depth of the
shower maximum at ultra-high energies could be explained by a very light
cosmic ray mass composition in combination with a
modification of the extrapolation of the hadronic cross section to
ultra-high energy.

\bibliographystyle{unsrt-mod-notitle}
\bibliography{RalfUlrich}

\end{document}